# Comparison of Pencil beam, Collapsed cone and Monte-Carlo algorithm in radiotherapy treatment planning for 6 MV photon

**Sung Jin Kim[†, ††], Dong Ho Kim[†], Sung Kyu Kim[†††].**

**Department of Physics, Yeungnam University, Gyeongsan. [†]**

**Department of Radiation Oncology, Eulji University Hospital, Daejeon. [††]**

**Department of Radiation Oncology, Yeungnam University College of Medicine, Daegu, Republic of Korea. [†††]**

# Comparison of Pencil beam, Collapsed cone and Monte-Carlo algorithm in radiotherapy treatment planning for 6 MV photon

**Abstract**

Purpose: Treatment planning system calculations in inhomogeneous regions may present significant inaccuracies due to loss of electronic equilibrium. In this study, three different dose calculation algorithms, pencil beam (PB), collapsed cone (CC), and Monte-Carlo (MC), provided by our planning system were compared to assess their impact on the three-dimensional planning of lung and breast cases.

Methods: A total of five breast and five lung cases were calculated using the PB, CC, and MC algorithms. Planning treatment volume (PTV) and organs at risk (OAR) delineation was performed according to our institution's protocols on the Oncentra MasterPlan image registration module, on 0.3 – 0.5 cm computed tomography (CT) slices taken under normal respiration conditions. Four intensity-modulated radiation therapy (IMRT) plans were calculated according to each algorithm for each patient. The plans were conducted on the Oncentra MasterPlan (PB and CC) and CMS Monaco (MC) treatment planning systems, for 6 MV. The plans were compared in terms of the dose distribution in target, OAR volumes, and monitor units (MUs). Furthermore, absolute dosimetry was measured using a three-dimensional diode array detector (ArcCHECK) to evaluate the dose differences in a homogeneous phantom.

Results: Comparing the PB, CC, and MC algorithms planned dose distributions, the PB algorithm provided adequate coverage of the PTV. The MUs calculated using the PB algorithm was less than those of the other algorithms. The MC algorithm showed the highest accuracy in terms of the absolute dosimetry.

Conclusion: Differences were found when comparing the calculation algorithms. The PB algorithm estimated higher doses for the target than the CC and MC algorithms. The PB algorithm actually overestimated the dose compared with those calculated by the CC and MC algorithms. The MC algorithm showed better accuracy than the other algorithms.

Keywords: dose calculation algorithm, collapsed cone algorithm, Monte-Carlo algorithm

## I. INTRODUCTION

In radiation therapy, the accuracy of dose calculations by a treatment planning system (TPS) is important to achieve tumor control and to spare normal tissue. An ideal dose calculation algorithm can perfectly reflect the actual dose distribution within a real patient, which in turn reduces the uncertainty during the evaluation of treatment plans. At present, the Monte-Carlo (MC) simulation is the most sophisticated and accurate algorithm [1-3].

As dose distribution tends to complicated in heterogeneous media, the dose calculation results show differences according to the algorithms. In accordance with the volume of heterogeneous media, this phenomenon clearly shows as it affects the absorption and scattering of beam. In case of the dose calculation with the PB algorithm on heterogeneous media which includes the low density region, the dose tends to be overestimated compared to the other algorithms as it hardly explains the phenomenon of the spread out electrons. In comparison, the MC algorithm remedies these problems, it ultimately demanded to use in clinical.

The problems of secondary build-up and electron transport phenomena in low-density regions have created the general impression that the use of MC methods is ultimately required for clinical dose calculations. Despite the proven accuracy of the MC method and the potential for improved dose distribution to influence treatment outcomes, the long calculation times previously associated with MC simulation rendered this method impractical for routine clinical treatment planning. However, the development of faster codes optimized for radiotherapy calculations and improvements in computer processor technology have substantially reduced calculation times to, in some instances, a matter of minutes on a single processor [4-6].

In the case of photon beams, the most modern calculation algorithms implemented in three-dimension planning systems are the pencil beam (PB) and super-position/convolution techniques, such as the collapsed cone (CC)[7-8]. The CC algorithm implements various approximations in the physics of radiation transport, which reduces the calculation time to levels that are acceptable for clinical practice. While PB algorithm is very fast, the limitations of PB algorithms in heterogeneous media are well known. This is because PB algorithms use a one-dimensional density correction which does not accurately model the accurately model the distribution of secondary electrons in media of different densities [9-10].

There are many theses which studied on algorithms such as comparison of the PB and CC, the PB and MC. Meanwhile, there are only few researches which compared and measured between 3 algorithms.

In this study, three different dose calculation algorithms, the PB, CC, and MC, provided by a commercial treatment planning system were compared to assess their impact on intensity-modulated radiation therapy (IMRT) of lung and breast cases. We will analyze effect of the treatment plan by the algorithm and consider the considerable part.

## II. MATERIALS AND METHODS

The plans were performed on computed tomography (CT) images of five lungs and five breasts using IMRT. Each case was calculated with the PB, CC, and MC algorithms. The plans were conducted on the Oncentra MasterPlan (PB and CC; V4.1, Nucletron, Veenendaal, NL) and MONACO (MC; V3.0, Elekta/CMS, Crawley, UK), for 6MV.

### Patients and plan information

All patient data were acquired from the same CT scanner with slice thicknesses of 0.3 to 0.5 cm under normal respiration conditions. Then, the information was transferred via Digital Imaging and Communications in Medicine (DICOM) to a TPS. After performing the patient contouring and localization of the treatment center, the treatment plans were computed for each case by each planning system using their respective dose calculation. The following cases were chosen:

(a) Breast: opposed beams with the medial field border aligned to avoid unnecessary irradiation of the underlying lung tissue (Fig 1(a)). The PTV includes the lumpectomy cavity yet let the intervals – 5mm each- between skin and thorax.

(b) Lung: a five-field technique with different weights on all beams to treat a gross tumor volume (GTV) in the right lung (Fig 1(b)).

More detailed information about the study cases is provided in Table 1. The established plans were delivered by an Elekta Synergy Platform (Elekta, Stockholm, Sweden). In all of treatment plan, the MLC was applied as Step and Shoot method.

### Dose calculation models

To compare the plans according to each algorithm, we used the PB convolution, the CC convolution and a Monte-Carlo calculation. The included TPSs in this study were from two of the vendors in the radiotherapy community, i.e., Nucletron for the Oncentra MasterPlan (OMP) and Elekta CMS for MONACO.

The OMP system has two different models, such as PB convolution/superposition and CC convolution. The models are based on energy fluence and also include head scatter modelling. The first model is based on a two-dimensional PB convolution for volume integration. Inhomogeneities are handled by an equivalent path length correction for the primary dose contribution and a one-dimensional convolution along fan lines with an exponential for scattered radiation [11-12]. The second model in the OMP is a CC convolution approach in which a ray-trace procedure through the irradiated object is utilized to get the TERMA at all points in the dose calculation matrix. The TERMA is separated into a primary part (collision kerma) and a scatter part, each of which are transported separately along 106 lines from the interaction point. The energy from each voxel intersected by a fan line in the irradiated medium is collected and deposited according to the elemental composition of the medium and density variations along the fan line [13-15].

The system from Elekta (CMS MONACO/MC) is based on a fluence model using a virtual energy fluence (VEF) model, while the dose distribution within the patient is calculated by the Photon Voxel MC algorithm XVMC [16-20].

**Plan comparison and evaluation tools**

Each of plans was then compared in terms of the dose distribution in target and the OAR volume. Dose-volume parameter, such as $V_{90\%}$ (the volume that is covered by 90% of prescription dose), $V_{95\%}$, $V_{100\%}$, $D_{5\%}$ (the dose that is irradiated by 5% of the volume of the planning treatment volume (PTV)), and $D_{95\%}$ for PTV, were used to compare the doses calculated by the PB, CC, and MC algorithms. For the lungs, the mean dose and the volumes receiving more than 5 Gy and 20 Gy were recorded. The mean dose and the volumes receiving more than 5 Gy and 30 Gy were recorded for the heart. Furthermore, the mean dose for the trachea and esophagus and the maximum dose for the spinal cord also were recorded.

Gamma analysis was also used to evaluate the dose distributions calculated by the PB, CC, and MC algorithms. The absolute dosimetry was measured using a three-dimensional diode array detector (ArcCHECK, Sun Nuclear, FL, USA).

**Statistical analysis**

Each values of PTV and OARs has checked for statistical differences through the Kruskal-Wallis test. For multiple comparisons, the Bonferroni compensation method was applied. The threshold for statistical significance was $P \leq 0.05$.

## III. RESULTS AND DISCUSSION

A summary of the PTV coverage is reported in Table 2. The values of PTV coverage for the breast were higher for the PB algorithm than for the MC algorithm. The values of $V_{95\%}$, $D_{5\%}$, and $D_{95\%}$ for the lung were higher for the PB algorithm than for the CC and MC algorithms. The MC algorithm had the highest values of $V_{90\%}$ and $V_{100\%}$ for lung. While the PB and CC algorithms showed acceptable dose coverage because optimization was used by these algorithms, the MC algorithms showed a significant lack of dose coverage in some lung tissue. In this way, the treatment plan with the MC was showed low dose in the lung- included region. Fig. 2 shows the dose distributions obtained with each algorithm in the axial and sagittal planes. The largest decrease in minimum PTV doses was observed mainly near the interface of the lung tissue. The dose calculated by the PB algorithm actually overestimated the dose in the lung region inside the fields compared with those calculated by the CC and MC algorithms. This behavior was common to all patients. In case of low density region was included into irradiation region, it shows the lack tendency of lines in the MC treatment plan. Our result is strongly in agreement with the findings of Borges et al. [21], in which the PB algorithm overestimated the dose in PTV in breast irradiation by 12% to 20% with respect to a commercial MC algorithm. S. Cilla et al. [22] also reported a similar result in which the PB algorithm significantly overestimated the dose in PTV, especially in the posterior part of the target facing the lung tissue, while the CC algorithm showed reduced target homogeneity, changed lung dose distribution, and a lower reference point dose.

Result on the OARs, such as the lung, heart, trachea, esophagus, and spinal cord, are summarized in Table 3. There were few differences in the doses for most organs. The mean dose and volume received dose of the lung and heart were higher in the PB algorithm for breast. For lung, the mean doses for the Rt. lung and heart were higher in the PB algorithm, while all the other values calculated by the MC algorithm were higher than those of the other algorithms.

Table 4 shows the motor units (MUs) for each algorithm. The MUs were higher in the order of PB, CC, and MC. The PB algorithm underestimated the MU required to achieve dose coverage. As the MC line evaluation was showed low in the low density region, it seems the MU of MC needs to be increased to give same amount of lines

The results of the three-dimensional gamma passing rate using ArcCHECK are shown in Table 5. The following criteria were applied to the distance to agreement-dose difference (DTA-DD) 3 mm/3%. Comparing the planned dose to the measured dose, the three-dimensional gamma passing rates were 91.75±9.12% under the PB algorithm, 93.12±7.75% under the CC algorithm, and 94.52±5.85% under the MC algorithm with 3 mm/3%.

## IV. CONCLUSION

In this study, we compared the dose algorithms calculated by PB, CC, and MC algorithms. Five patients with breast and five patients with lung cancer were quantitatively analyzed for PTV volume and OARs. The volume of the PTV covered by 95% of the prescription dose was superior for the PB algorithm. The PB algorithm overestimated the dose and, thus, underestimated the MU required to achieve dose coverage. A comparison of the results of the PB, CC, and MC algorithms using ArcCHECK showed that a more accurate gamma passing rate was obtained when the MC algorithm was used.

If previous clinical experiences are based on the use of the PB algorithm, one needs to fully understand the dosimetric changes.

I don’t insist the excellence of one algorithm in this thesis. There are many difficulties to generalize the problem; such as experimental groups, methods. However, considering the each algorithm’s attribute while establish the treatment plan, it will help to realize the optimum treatment plan.

Table 1. Planning information about all study cases used for the evaluation.

Table 2. Comparison of dosimetric data for PTV coverage. (± standard deviation)

Table 3. Comparison of dosimetric data for OARs. (± standard deviation)

Table 4. The MUs for each algorithm. (± standard deviation)

Table 5. The results of 3D gamma passing rate using the ArcCHECK. The criteria applied DTA-DD were 3mm-3%. (± standard deviation)

Table 1. Planning information about all study cases used for the evaluation.

| Case | The number of patients | Prescription dose | Constraint of OARs | Dose fraction | Beam arrangements |
|---|---|---|---|---|---|
| Breast | 5 | PTV 50 Gy | Lt. lung (*20%vol./20Gy) Heart (10%vo.l/30Gy) | 25 | opposed beams |
| Lung | 5 | PTV 66Gy | Lung (20%vol./20Gy) Heart (10%vol./30Gy) Spinal cord (max. 45Gy) | 33 | G=10°, 180°, 225°, 270°, 315° |

* x vol. / y Gy : The constraint of dose (y) to the volume of OARs(x).

Table 2. Comparison of dosimetric data for PTV coverage (± standard deviation)

| | | PB | CC | MC | P < 0.05* |
|---|---|---|---|---|---|
| Breast | $^{a}V_{90\%}$(%) | 98.03±1.47 | 98.03±1.59 | 94.69±1.64 | e2, e3 |
| | $^{a}V_{95\%}$(%) | 94.41±2.82 | 93.53±3.19 | 85.55±3.21 | e2, e3 |
| | $^{a}V_{100\%}$(%) | 46.18±5.80 | 48.46±5.34 | 25.54±10.29 | e2, e3 |
| | $^{b}D_{5\%}$(Gy) | 47.22±1.05 | 46.99±0.99 | 44.14±2.03 | e2, e3 |
| | $^{b}D_{95\%}$(Gy) | 45.25±1.80 | 45.15±17.69 | 36.14±6.12 | e2, e3 |
| Lung | $^{a}V_{90\%}$(%) | 99.90±0.12 | 99.74±0.24 | 99.92±0.16 | - |
| | $^{a}V_{95\%}$(%) | 99.40±0.31 | 98.07±1.34 | 98.97±1.24 | - |
| | $^{a}V_{100\%}$(%) | 63.77±5.26 | 59.50±2.77 | 76.15±11.83 | e2 |
| | $^{b}D_{5\%}$(Gy) | 64.90±0.29 | 64.22±0.73 | 64.65±0.86 | e2, e3 |
| | $^{b}D_{95\%}$(Gy) | 64.06±0.46 | 62.95±1.08 | 63.74±1.18 | - |

$^{a}V_{x\%}$ : The volume for x percent of prescription dose.

$^{b}D_{y\%}$ : The dose for irradiated by y percent of the volume of the PTV.

* Bonferroni analysis : e1 = PB vs. CC; e2 = CC vs. MC; e3 = MC vs. PB.

Table 3. Comparison of dosimetric data for OARs (± standard deviation)

| | | | PB | CC | MC | P < 0.05* |
|---|---|---|---|---|---|---|
| Breast | Ipsilateral lung | $^{b}$Mean(Gy) | 11.92±0.65 | 11.36±0.72 | 10.06±0.40 | e2, e3 |
| | | $^{a}V_{5Gy}$(%) | 32.74±3.81 | 35.24±3.94 | 31.54±3.17 | - |
| | | $^{a}V_{20Gy}$(%) | 22.64±1.01 | 22.63±1.08 | 20.05±0.36 | - |
| | Heart | $^{b}$Mean(Gy) | 12.48±1.70 | 11.98±1.78 | 8.21±0.68 | e2, e3 |
| | | $^{a}V_{5Gy}$(%) | 44.14±5.91 | 41.61±6.97 | 30.05±3.49 | e2, e3 |
| | | $^{a}V_{30Gy}$(%) | 16.21±2.84 | 15.88±2.84 | 9.78±0.48 | e2, e3 |
| Lung | Rt.lung | $^{b}$Mean(Gy) | 28.94±12.79 | 29.21±12.54 | 28.83±11.60 | - |
| | | $^{a}V_{5Gy}$(%) | 71.77±21.61 | 75.71±19.88 | 76.65±20.05 | - |
| | | $^{*}V_{20Gy}$(%) | 58.57±23.75 | 60.93±22.63 | 61.17±22.75 | - |
| | Lt.lung | $^{b}$Mean(Gy) | 9.34±2.23 | 9.42±2.00 | 9.52±1.91 | - |
| | | $^{a}V_{5Gy}$(%) | 70.75±13.67 | 72.00±13.51 | 72.65±12.33 | - |
| | | $^{a}V_{20Gy}$(%) | 6.16±4.81 | 5.87±3.97 | 6.57±6.57 | - |
| | Heart | $^{b}$Mean(Gy) | 9.90±7.69 | 9.65±7.49 | 9.08±6.50 | - |
| | | $^{a}V_{5Gy}$(%) | 51.97±40.74 | 50.63±40.66 | 52.48±39.34 | - |
| | | $^{a}V_{30Gy}$(%) | 5.41±6.34 | 4.88±5.62 | 3.12±3.85 | - |
| | Trachea | $^{b}$Mean(Gy) | 22.16±9.88 | 22.39±9.28 | 23.62±9.06 | - |
| | Esophagus | $^{b}$Mean(Gy) | 23.45±10.71 | 23.98±10.30 | 23.46±9.62 | - |
| | Spinal cord | $^{c}$Max(Gy) | 46.56±3.29 | 45.62±3.87 | 42.65±1.18 | - |

[a] $V_{x\%}$ : The volume for x percent of prescription dose.

[b]Mean : The mean dose for OARs.

[c]Max : The Maximum dose for OARs.

* Bonferroni analysis : e1 = PB vs. CC; e2 = CC vs. MC; e3 = MC vs. PB.

Table 4. The MUs for each algorithm. (± standard deviation)

| | algorithms | | |
|---|---|---|---|
| | PB | CC | MC |
| Breast | 293.82±51.51 | 298.34±42.78 | 346.76±55.46 |
| Lung | 321.68±34.14 | 350.44±42.73 | 412.21±49.84 |

Table 5. The results of the three-dimensional gamma passing rate using ArcCHECK. The criteria applied to the DTA-DD were 3 mm/3% (± standard deviation)

| Calculation algorithm | 3D gamma passing rate (%) |
|---|---|
| | 3mm-3% |
| PBC | 91.75±9.12 |
| CCC | 93.12±7.75 |
| MCC | 94.52±5.85 |

Fig. 1. Images of a patient slice with the beam arrangement for (a) breast cases, and (b) lung cases

Fig. 2. Dose distributions obtained with each algorithm in the axial and sagittal planes for (a)

breast cases, and (b) lung cases

Fig 1. Images of a patient slice with the beam arrangement for (a) breast cases, and (b) lung cases.

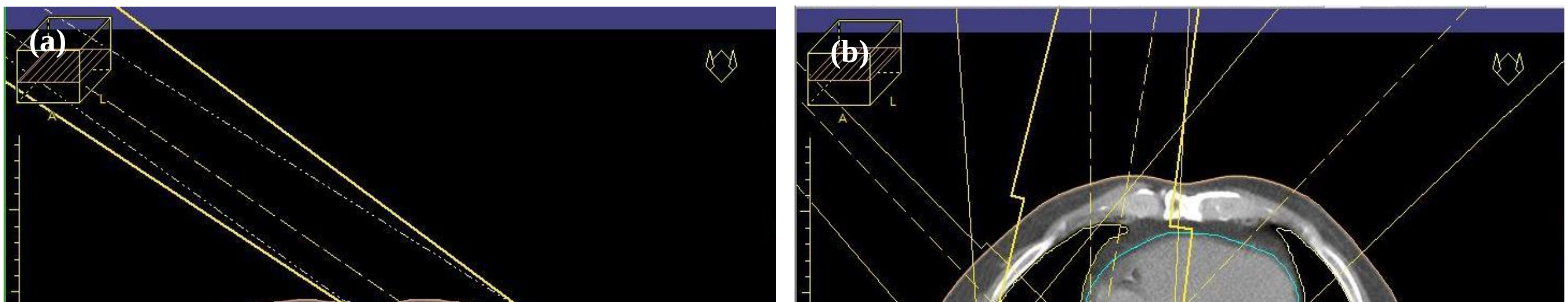

Fig 2. Dose distributions obtained with each algorithm in the axial and sagittal planes for (a) breast cases, and (b) lung cases.

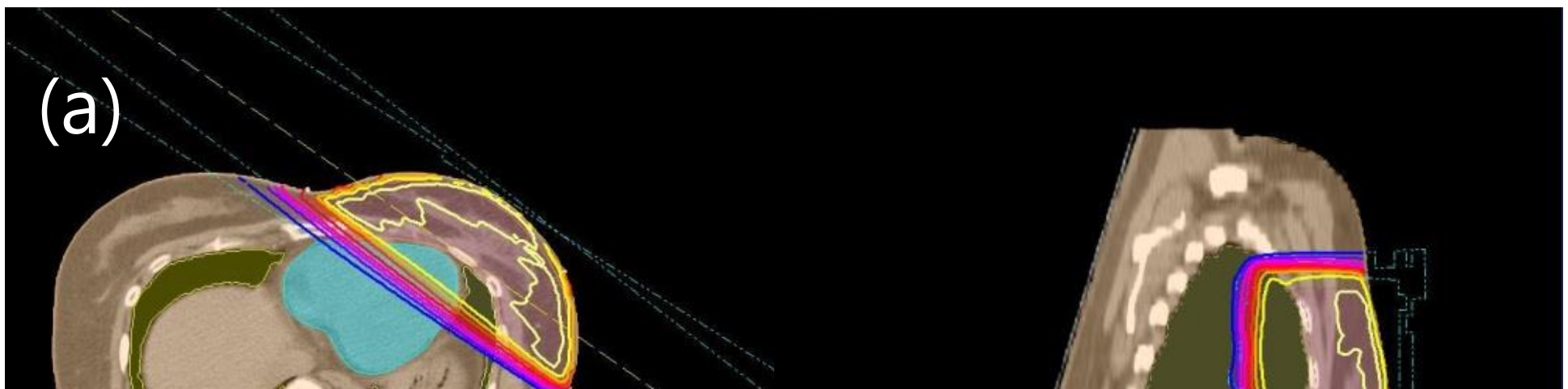

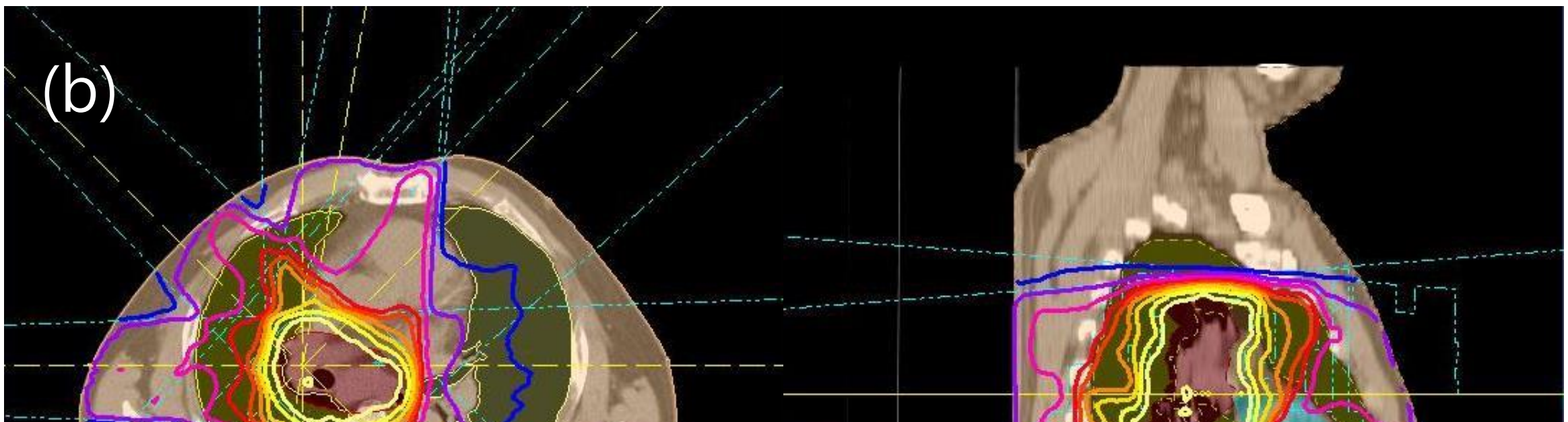
(b)